# Buffer-Enhanced Electrical-Pulse-Induced-Resistive Memory Effect in Thin Film Perovskites

Xin CHEN,* Naijuan WU, Alex IGNATIEV

*Texas Center for Advanced Materials, University of Houston, Houston, TX 77204, USA*

Qing CHEN, Yue ZHANG

*Key Laboratory for the Physics and Chemistry of Nanodevices, Department of Electronics, Peking University, Beijing 100871, P. R. China*

A multilayer perovskite thin film resistive memory device has been developed comprised of: a $Pr_{0.7}Ca_{0.3}MnO_3$ (PCMO) perovskite oxide epitaxial layer on a YBCO bottom thin film electrode; a thin yttria stabilized zirconia (YSZ) buffer layer grown on the PCMO layer, and a gold thin film top electrode. The multi-layer thin film lattice structure has been characterized by XRD and TEM analyses showing a high quality heterostructure. I-AFM analysis indicated nano granular conductivity distributed uniformly throughout the PCMO film surface. With the addition of the YSZ buffer layer, the pulse voltage needed to switch the device is significantly reduced and the resistance-switching ratio is increased compared to a non-buffered resistance memory device, which is very important for the device fabrication. The magnetic field effect on the multilayer structure resistance at various temperatures shows CMR behavior for both high and low resistance states implying a bulk material component to the switch behavior.

* E–mail address: xinchen@svec.UH.EDU

**Introduction**

The recent discovery of the electrical pulse induced resistance change effect (EPIR)[1)] in perovskite oxide thin films at room temperature has drawn wide attention[2-4)]. A pervoskite oxide film such as $Pr_{0.7}Ca_{0.3}MnO_3$ (PCMO) sandwiched between two electrodes forms a two-terminal EPIR device. The resistance of the device can be reduced by an electric pulse of one polarity and increased by a pulse of the opposite polarity. Such reversible resistance switching devices have important application potential in future non-volatile high density resistive random access memory devices (RRAM).[5)] The resistance-switching of EPIR devices based on a PCMO film, a colossal magnetoresistance (CMR) material, has been repeated by several groups at room temperature. However, the physical origin of the resistance switching is not yet fully elucidated. Both bulk[6)] and interface effects[7)] have been proposed for the EPIR mechanism in the CMR perovskites. The colossal magnetoresistance (CMR) nature of the PCMO material brings properties such as percolations, charge/orbital ordering, nanometer phase separation, lattice structure distortion, composition deviations, interface trapping states, and thin film morphology into consideration in the definition of the mechanism.[8-11)] Considering the reported importance of the metal-oxide interface in the device switching property,[7)] we report here our effort to modify the interface in an EPIR device.

In this paper, we report on a four-layered CMR perovskite-based EPIR device, where a thin insulating yttria stabilized zirconia (YSZ) film is used as a buffer layer to modify the interface properties between the PCMO thin film and the top electrode with a resulting positive effect. In addition, we obtain further information on bulk vs. interface mechanisms through the study of the relationship between resistance switching properties and the lattice micro structure of the device layers, and the changes of high- and low-resistance states under applied magnetic field.

**Experimental and discussion**

Fig. 1(a) shows the resistance change of a traditional EPIR device.[1)] A $Pr_{0.7}Ca_{0.3}MnO_3$ (PCMO) film of 600 nm thickness was epitaxially deposited by pulsed laser deposition (PLD) on top of a 300 nm YBCO bottom electrode layer previously grown on an $LaAlO_3$ (100) (LAO) substrate. The structure was then covered with a gold top electrode. Short electrical pulses were then applied across the top gold electrode and the bottom YBCO electrode with the device resistance change after 200ns single pulses shown in Fig. 1(a). The low resistance ($R_L$) of ~250Ω was obtained after a +13V pulse was applied, and the high resistance ($R_H$) of ~400Ω was obtained after a -13V pulse was applied. The resistance of the sample was measured with a very small sensing current of ~1μA. The positive pulse direction is defined as from the top electrode to the bottom electrode. The switch ratio $(R_H-R_L)/R_L$ in this sample was ~60%.

Fig. 1(b) shows the resistance change versus electrical pulse number for the new PCMO-based four-layer buffer-EPIR device where a YSZ buffer layer has been introduced between the PCMO layer and the Au top electrode. A YSZ film about 20nm thick was deposited by PLD on the PCMO/YBCO thin film heterostructure grown on the LAO (100) substrate, which was fabricated at the same time as the sample used in Fig. 1(a). The YSZ deposition was performed at 650°C, which allowed for good surface coverage of the film without a YSZ/PCMO reaction. Fig. 1(b) shows that the device resistance increased significantly with the addition of the insulating YSZ buffer layer, while the voltage needed for switching the device was significantly reduced to ~ ±3V in comparison with the ~ ±13V of Fig. 1(a). In addition, the switch ratio was increased to ~70% as shown in Fig. 1(b). Such advantages of a buffer layer integrated four-layer EPIR device have been confirmed in repeated experiments.

Fig. 1(c) is the non-volatile resistance hysteresis measurement for the buffered Au/YSZ/PCMO/YBCO sample, and shows a simple loop similar to the hysteresis curves we have obtained for non-buffer layer PCMO-based EPIR devices with conducting oxide bottom electrodes.[12)] Fig. 1(c) shows a rapid transition to the high resistance ($R_H$) or to the low resistance ($R_L$) state that is indicated by the nearly rectangular hysteresis loop. The $R_H$ and $R_L$ states obtained in Fig. 1(c) are not exactly at the same value as in Fig. 1(b) due to the multiple pulses applied to the device in the measurements of Fig. 1(c).

It should be noted that the switching ratio of the buffer-EPIR device can be made much larger (>1000%) by the application of higher pulse voltage. We have concentrated on the low voltage regime first to protect the sample from pulse damage, and secondarily because it is very important for the device industry to have devices working under 5V. We need to point out that although there are ways to improve an EPIR device, the standard EPIR device shown in Fig. 1(a), exhibited almost no switching under ±5V pulses.

Pulse duration dependence of the switching of buffer-EPIR device samples has been studied and as shown in Fig. 1(d), lower voltages can be used to switch the device under longer pulses, indicating that the switching voltage may not reflect a response due to overcoming some physical barrier height in the switching process, but that the total energy added to the device, or the total charge (through current) may be drivers for the resistive switching process. .

The possibility that the resistance switching behavior observed in the buffered device might be due to the resistance switching of the YSZ buffer layer itself was addressed by monitoring the resistance switching properties of a Au/YSZ/YBCO structure, i.e. a structure without the PCMO pervoskite layer. For this test, a 20nm YSZ film was grown on a YBCO/LAO substrate under the same conditions as for the sample of Fig. 1(b). The results shown in Fig. 2 indicate that the resistance of the YSZ film on YBCO can be partially switched, but the switch ratio rapidly decays. In addition, the Au/YSZ/YBCO sample requires a much higher switching pulse voltage of ±7.3V as compared to the ~3V switch voltage needed for the Au/YSZ/PCMO/YBCO sample of Fig. 1(b). The switching of the YSZ device is unanticipated and indicates that there may be a switching mechanism common to many oxides. The instability of the switching with the exhibited rapid decay may be due to structural defects and resultant 'burn through' in the very thin 20 nm YSZ film.

Fig. 3 shows the magnetic field dependence of the four-layer buffer-EPIR device resistance as a function of temperature. Under no applied magnetic field, the device resistance increased rapidly when temperature was decreased for both the high and low resistance states. Under application of an 8 Tesla magnetic field, the device resistance showed CMR behavior at low temperature with an R-T hysteresis loop in the temperature range of 80K to 120K. The temperature-rate of change at room temperature of the four-layer device can be related to the activation energy for the decerase of resistnace with temperature. A value of Ea~130meV was obtained from the data in Fig. 3, which is similar to the value for manganites[13], and much lower than that for bulk YSZ which has a value of Ea~1eV[14]. This indicates that the behavior of resistance with temperature of the buffer-EPIR device is principally given by the PCMO manganite property of the device, and not by the YSZ layer due to its small thickness and low mass fraction.

Two points should be pointed out concerning the resistance change measurements of the buffer-EPIR device under 8T magnetic field: (1) At room temperature, both $R_H$ and $R_L$ are reduced by ~3% as compared to no applied magnetic field. In addition, with decreasing temperature, the resistance change of $R_H$ and $R_L$ under 8T magnetic field are different as seen in the insert of Fig. 3. This dependence of $R_L$ and $R_H$ on magnetic field indicates that the PCMO CMR film is playing a role in the resistance switching process. (2) Upon cooling under applied 8T magnetic field, an insulator-metal transition is seen at ~90K. On heating from 60K, the metal-insulator transition occurs at ~120K, showing a hysteretic behavior. As part of this hysteretic behavior, it is seen that the transition under cooling was by a two-step transition process. Similar behavior has recently been observed in a CMR $La_{1-x}Ca_xMnO_3$ material.[15] It has been suggested[3] that there is a second minimum energy state in the CMR material at low temperature, where the electron spins of neighboring atoms are neither parallel nor anti-parallel, but with a certain angle $\theta$. This might explain the two-step insulator-metal transition observed. It is also known that in the high-temperature regime in CMR materials, small magnetic correlations and electric charge/lattice interplay both exist,[13] hence the small, but clear, magnetic dependence observed in both $R_H$ and $R_L$ states at room temperature can be considered additional evidence that the PCMO film plays a role in the resistance behavior of the EPIR device.

The lattice micro structure of the four-layer buffer-EPIR device was examined by transmission electron microscopy (TEM) using a FEI Tecnai F30 microscope in an attempt to identify any interim film morphology and help to further illustrate the switching mechanism.

Fig. 4 shows cross-sectional low magnification and high resolution TEM images of the interfaces in Au/YSZ/PCMO/YBCO/LAO sample. Fig. 4(a) is a low magnification micrograph indicating a 300nm thick YBCO layer on LAO and a 660 nm thick PCMO layer on the YBCO layer. The PCMO layer has column-like structure starting from the PCOM/YBCO interface, with the column axis parallel to the film normal and a column width of a few tens of nanometers. The interface between LAO and YBCO and that between YBCO and PCMO are flat, though some strain exists in these two interfaces as indicated by the image contrast at the interface. Our study shows that the YBCO layer epitaxially grown on LAO (100) has high crystalline quality with c-axis perpendicular to the LAO (100) surface. The high resolution TEM (HRTEM) image of the PCMO/YBCO cross section (Fig. 4(b)) shows that the PCMO layer is epitaxially grown on the c-YBCO film. Fast Fourier transform (FFT) patterns taken from the YBCO area and from the PCMO area indicate that the $(10\bar{1})$-oriented PCMO was grown with two in-plan aligned domains: $[010]_{PCMO}$//$[010]_{YBCO}$ (3.84Å vs. 3.88 Å) (shown in Fig.4(b)), and $[101]_{PCMO}$ //$[010]_{YBCO}$ (3.86Å vs. 3.88 Å). Due to the slight difference between the PCMO and the YBCO lattice parameter in the a and b directions, the column-like structure seen in the PCMO layer may partially result from strain at the PCMO/YBCO interface and also partially from the two domain epitaxial growth of the PCMO thin film on YBCO. XRD θ-2θ and GADDS spectra concur with the epitaxial orientation of YBCO and PCMO obtained by TEM. Furthermore, there is no observation of inter-diffusion between the PCMO and YBCO layers. The column boundaries in PCMO are crystalline but with strain probably due to lattice mismatch and structural defects.

The HRTEM image of Fig. 4(c) shows the microstructure of the YSZ nano layer and the interface between the YSZ/PCMO layers. It can be seen that the surface of the PCMO is not flat. The 10-25 nm thick YSZ layer is polycrystalline, containing randomly distributed domains with the domain sizes of about 5-20 nm. This is principally due to the fact that the YSZ was deposited at moderate temperature (650°C) to minimize any YSZ/PCMO inter-diffusion. Thus inter-diffusion between these two layers was not observed and the interface between PCMO and YSZ remained crystalline. Further analysis shows that the thin YSZ layer growth largely followed the topography of the PCMO, and completely covered the nano-corrugated surface of the PCMO, although the YSZ thickness on the PCMO is not very uniform.

Further measurement of the buffer-EPIR and standard EPIR device samples were made via a current imaging I-AFM (using a PSIA XE-100 SPM). This allowed for measurement of the conductivity distribution in the surface of the PCMO layer in the PCMO/YBCO/LAO structured EPIR device. Fig. 4(d) is the AFM plot showing the nanogranular structure of the PCMO surface in correlation with the column-like morphology of the PCMO shown by TEM in Fig. 4(c). Fig. 4(e) is the I-AFM plot which shows that the nano grains of Fig. 4(d) are highly conducting regions in the PCMO surface (bright) with low conducting boundaries between the grains (dark). This suggests that current mainly passes along the columnar grains of our PCMO films and quite uniformly over the whole sample instead of the filamentary conduction as reported in the other oxides such as Cr-doped $SrZrO_3$.[16] This I-AFM measurement is made on a virgin sample, because it can only be tested on a bare surface instead of in a switched film under the electrode layer. Further research is underway on switched devices.

I-AFM measurements were also made on a YSZ/PCMO/YBCO/LAO sample surface. The I-AFM image showed the surface as a uniform dark area without major features relating to current flow. For this (and most) buffer layer sample, the insulating YSZ layer increased the device resistance by ~30 times, as shown in Fig. 1. Furthermore, no high current spots were observed on the YSZ surface further negating the proposed existence of filamentary paths[16] in the YSZ buffer layer that could lead to the switching phenomenon in EPIR devices.

The function of the YSZ buffer layer in improving the device switch properties can be considered in two ways: electronic contribution and ionic contribution. Charge injection/extraction and accumulation has been suggested as playing a role in the EPIR effect, especially at the interface.[8] The charge carriers in the PCMO near the interface between the insulating YSZ layer and the semiconducting PCMO layer (the PCMO-interface) could be accumulated and depleted during the application of positive and negative electric pulses. This accumulation/depletion could be modified by the inclusion of the YSZ buffer layer which could cause a larger carrier density change in the PCMO-interface as compared to the change throughout the CMR material without an insulating buffer. The carrier density change would affect the number of carriers and sites available for hopping and would change the device resistance.

With respect to ionic contribution, YSZ is an oxygen ion conductor, and hence could enhance any movement of oxygen ions or oxygen vacancies within the interface region. Such movement could change interface resistivity, and would of course also affect any electronic charge build up at the interface. Additional work is underway to further clarify whether electronic and/or ionic conduction plays the major role in the EPIR switching process.

**Summary**

The addition of a thin insulating YSZ buffer layer between the top electrode and the active oxide in an EPIR device has shown improved switching behavior of the device including reduced pulse voltage and higher switching ratio. I-AFM measurements have shown no filamentary behavior in our device, and low temperature observation of the CMR behavior in the buffer-EPIR device for both the $R_L$ and $R_H$ states indicates that the PCMO layer is playing a role in the EPIR resistance change effect. Such enhancements may come from a change in carrier density in the PCMO layer near the YSZ interface, and/or from a change in oxygen ion/vacancy concentration near the interface. Additional efforts are under way to further elucidate resistance switching in perovskite films, as well as the development of resistive switching devices based on these materials.

**Acknowledgements**

The help from J. Strozier, Z.J. Tang, L.M. Peng, A.M. Guloy and A.J. Jacobson is very much appreciated. This research was partially supported by NASA, the State of Texas through TcSAM, Sharp Laboratories of America, and the R. A. Welch Foundation. This work was also supported by National Science Foundation of China Gant Nos. 60271004 and 60440420450.

Fig. 1(a) The resistance change of a PCMO device without buffer layer (b) Resistance change measurement of the PCMO device with a YSZ buffer layer (c) Hysteresis measurement of the PCMO device with a YSZ buffer layer (d) Pulse width dependence of switching on the magnetic test sample.

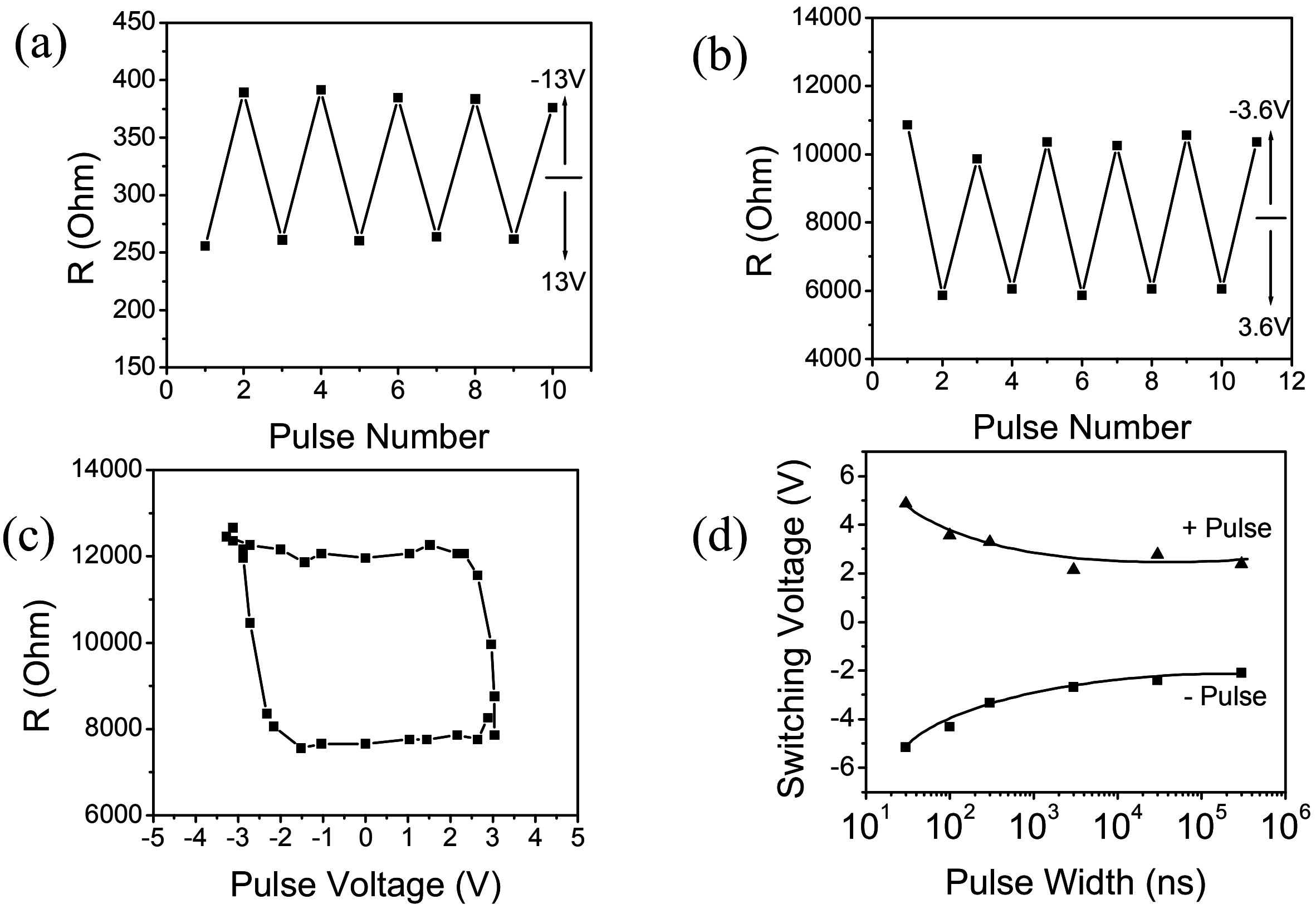

Fig. 2 Resistance change measurement of a YSZ layer with out the underlying PCMO film.

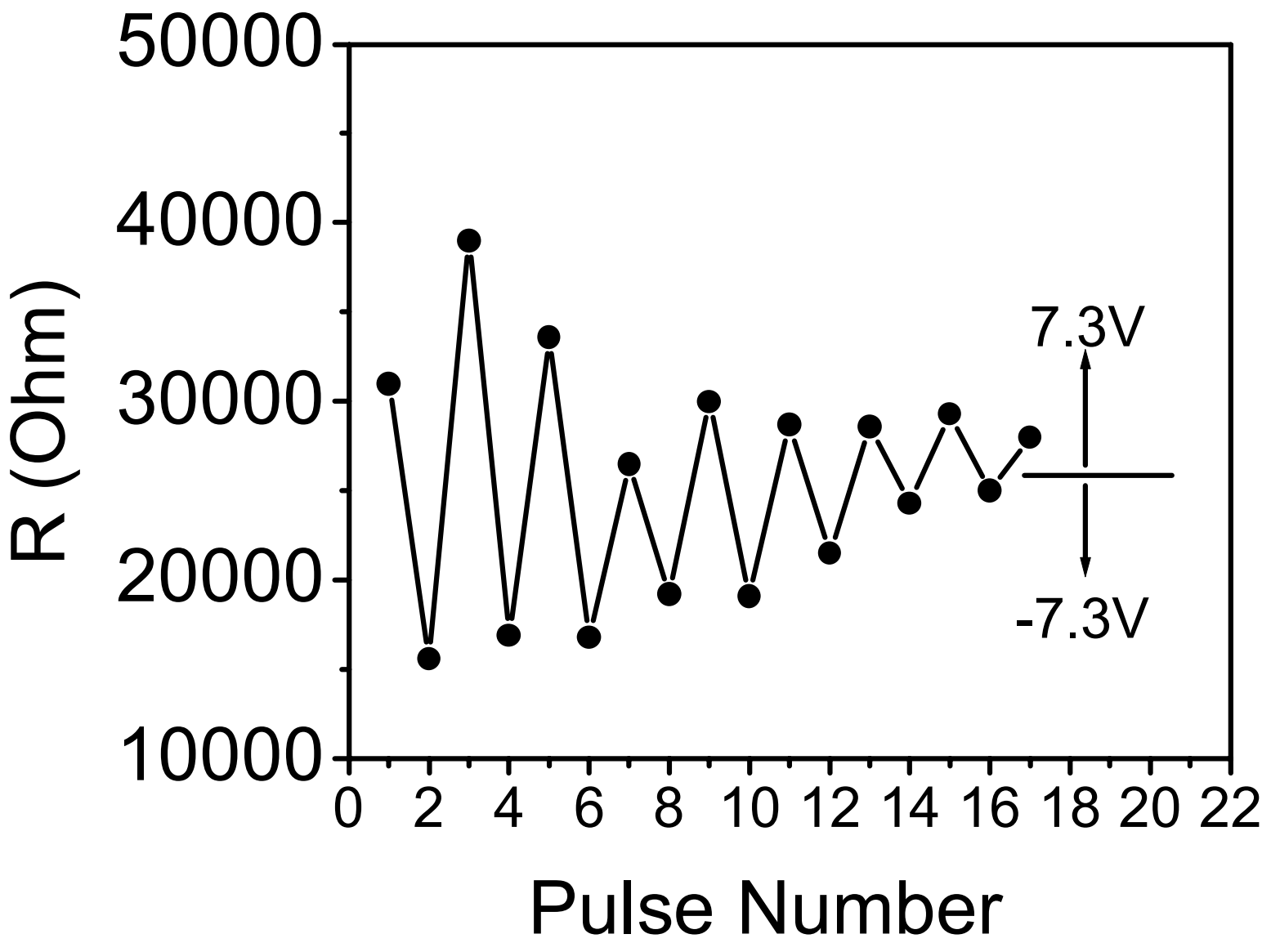

Fig. 3 Temperature dependence of a buffer-EPIR device at zero and 8 Tesla magnetic field (MR ratio = 3% at RT).

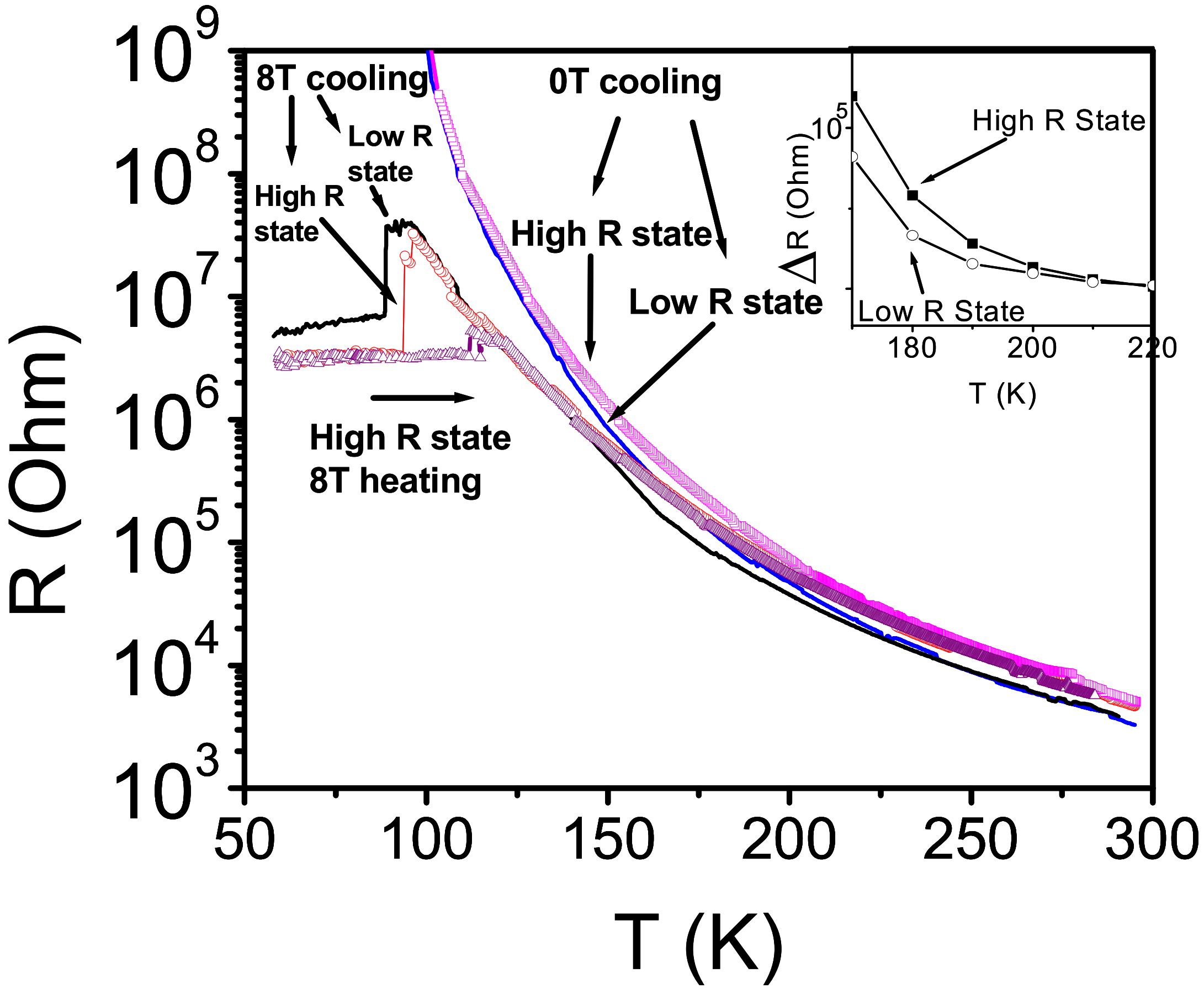

Fig.4 (a) TEM image showing the cross section view of the multi-layers structure (b) HRTEM image showing the interface between YBCO (top left) and PCMO (bottom right). The FFT patterns included are taken from the two area and showing the orientation of the two layers (c) A higher magnification image showing the top YSZ layer (d) AFM (e) I-AFM images on a PCMO/LAO sample surface.

(a) (b) (c) (d) (e)

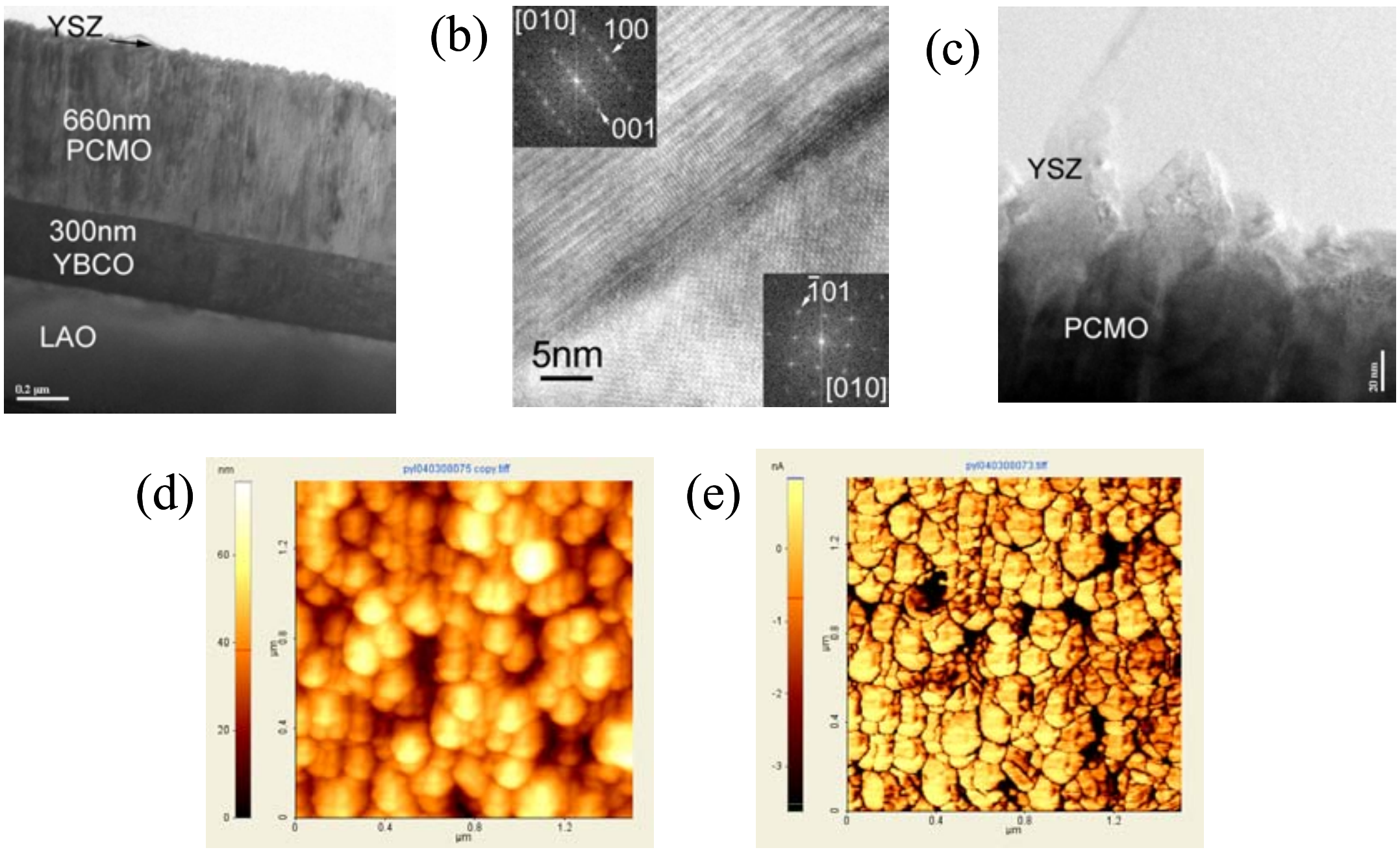